\documentclass[12pt,preprint]{aastex}

\usepackage{amsmath}

\shorttitle{PSR B0826-34: Sometimes a RRAT}
\shortauthors{Esamdin et al.}

\begin{document}

\title{PSR B0826-34: Sometimes a Rotating Radio Transient}
\author{
A. Esamdin,$^{1}$ D. Abdurixit,$^{2}$ R. N. Manchester,$^{3}$ H. B. Niu,$^{1}$
 }
\affil{$^{1}$Xinjiang Astronomical Observatory, Chinese Academy of Sciences, Urumqi 830011,
China; aliyi@xao.ac.cn}
\affil{$^{2}$Department of Physics, Xinjiang University, Urumqi 830046, China}
\affil{$^{3}$CSIRO Astronomy and Space Science, PO Box 76, Epping NSW 1710, Australia}
\affil{$^{*}$E-mail: aliyi@xao.ac.cn}

\begin{abstract}
  We report on the detection of sporadic, strong single pulses
  co-existing with a periodic weak emission in the duration of weak
  mode of PSR B0826-34. The intensities and durations of these pulses
  are comparable with that of the sub-pulses in the strong mode, and these
  pulses are distributed within the phase ranges of the main-pulse and interpulse
  of the strong-mode average profile. These results suggest that there are most possibly sporadic,
  very short timescale turn-on of strong-mode emission during the weak-mode
  state of the pulsar. The emission features of the bursts of strong pulses of
  PSR B0826-34 during its weak-mode state are similar to those of the
  rotating radio transients (RRATs). PSR B0826-34 is the second
  pulsar known which oscillates between pulsar-like and RRAT-like modes.
\end{abstract}


\section{Introduction}

PSR B0826-34 is a relatively old pulsar with a characteristic age of
$3\times10^{7}$ yr \citep{mlt+78}. An early study of this
pulsar at 408 MHz by  \citet{dll+79} showed that the pulsar exhibits nulls
for 70\% of the time. \citet{bmh+85} confirmed the large null fraction,
and showed that the emission of the pulsar at 645 MHz extends throughout the
pulse period (P=1.848 s) during its radiation state. They found bands
of drifting sub-pulses with a wide variation of drift rate, including reversal
of drift direction. Further investigation by \citet{elg+05} at 1374 MHz
detected a weak emission profile during the long null states of the pulsar, and they
argued that the null state was a weak-emission mode
rather than being a null. However, \citet{bgg+08} reported non-detection
of emission in a long null state of this pulsar by using the GMRT
at 157, 303, 325, 610 and 1060 MHz. Recent Parkes observations of the pulsar
at 685 and 3094 MHz confirm the existence of the weak mode emission \citep{ser+11}.
\citet{vlt+12} also argue the presence of the weak mode of the pulsar,
and suggest that the weak mode is a magnetospheric state that is different from
the strong mode.

We study the weak mode emission of PSR B0826-34 on an individual-pulse
basis by re-analyzing the data used by \citet{elg+05}. In this paper,
we report on the detection of the sporadic strong
pulses during the weak mode of the pulsar. The detection of single
pulses and their properties are described in Section 2. In Section 3,
an analysis of periodicity of the weak mode emission is presented. The
results are discussed in Section 4, and the conclusions are given in
Section 5.

\section{Sporadic strong pulses detected in the weak-mode state}

The data we analyzed were obtained on 2002 September 10 using the 64-m
Parkes radio telescope. The central beam of the multibeam receiver was
used at a central observing frequency of 1374 MHz with a $96\times
3$~MHz analogue filterbank \citep{mlc+01}. The data, lasting 359
minutes, were sampled at 2-ms intervals.

\begin{figure}[h,t]
\centerline{\includegraphics[angle=-90,width=0.85\textwidth]{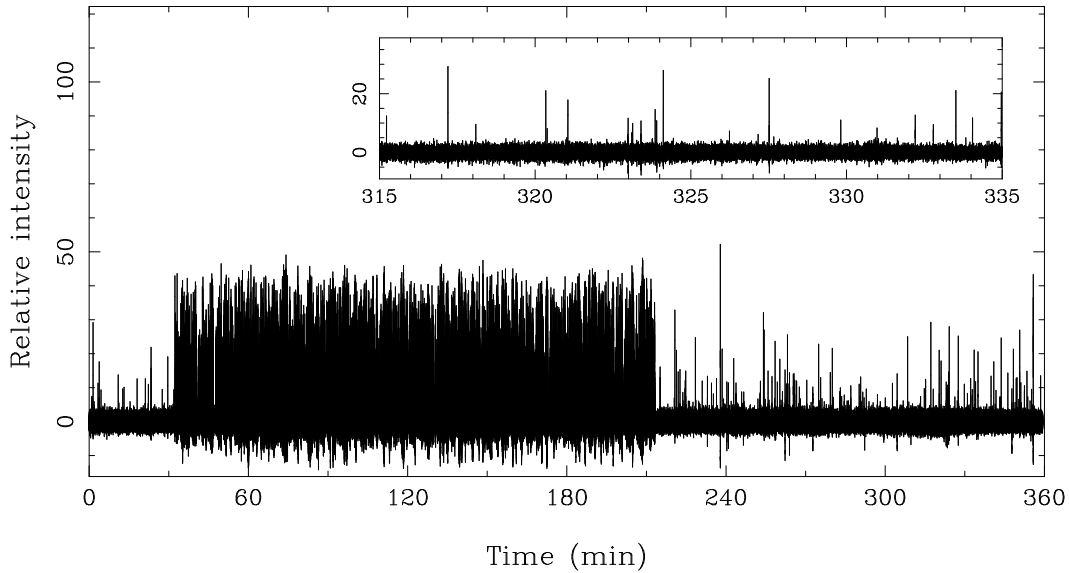}}
\caption{Pulse time series for PSR B0826-34. A block of continuous
  bursts of strong pulses is preceded and followed by long periods of
  weak-mode emission. A total of 153 strong pulses is detected in the
  weak-mode states. A section of 20 minutes of weak-mode state is
  presented in inner box, to show more clearly the sporadic bursts of
  the strong pulses. The negative-going features associated with strong
  pulses result from instrumental saturation and are not real. The
  units of intensity are arbitrary.}
\label{fg:ts}
\end{figure}

Figure~\ref{fg:ts} presents the time series of the data which
comprises 11663 individual pulses.  The emission in
Figure~\ref{fg:ts} exhibits a block of bursts of continuous strong
pulses preceded and followed by long periods of the weak-mode emission
which we describe as the weak-mode state hereafter.  As shown in
Figure~\ref{fg:ts}, the block of bright pulses present from 32 min to
213 min (5866 rotations).  This block of strong emission is
interrupted by 82 short sequences of weak-mode pulses, which last from
few to tens of the pulse periods. The weak-mode state, from 0 to 32
min and from 213 to 359 minutes in Figure~\ref{fg:ts},
includes 5797 pulse periods in total.  In this work, the two durations
of weak-mode state are investigated.

The outstanding feature of the two weak-mode blocks in
Figure~\ref{fg:ts} is the existence of rare, short-duration, strong
pulses. The inner box in Figure~\ref{fg:ts} presents 20 minutes of
weak-mode data, showing more clearly the rare single pulses. In 178
min of weak-mode state, 153 strong pulses with signal-to-noise ratios
(S/N) above the $5\sigma$ threshold are detected. The S/N is taken to
be the ratio of the peak amplitude in a given pulse period divided by
the RMS scatter of the points in the same period. The detection rate
of the strong pulses is about 51.4 per hour.

\begin{figure}[h,t]
\centerline{\includegraphics[angle=-90,width=0.65\textwidth]{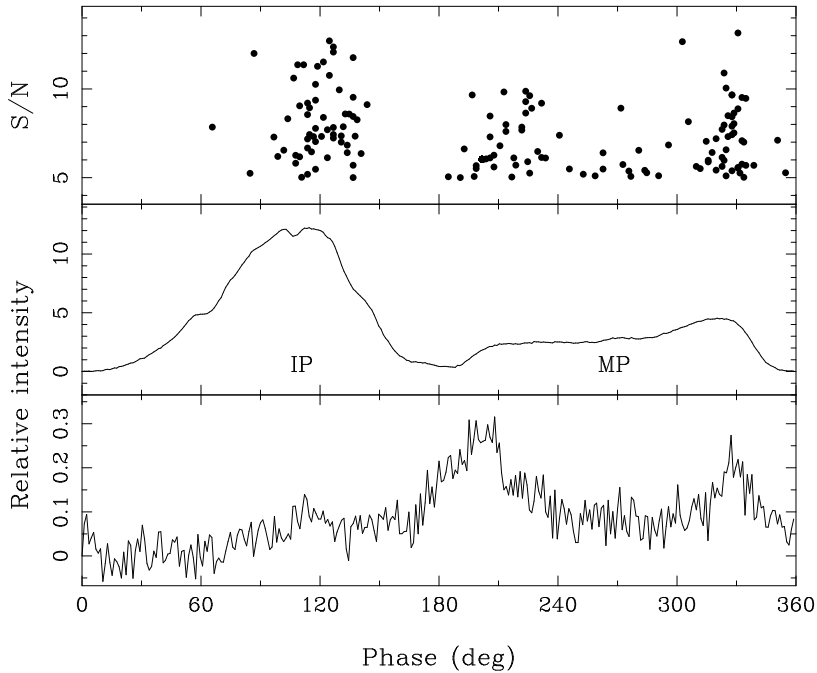}}
\caption{Phase-S/N distribution of the 153 detected strong
pulses (top panel) compared with the average profiles of the strong
(middle panel) and weak mode (bottom panel). Each mean profile was obtained
by folding 4500 successive individual pulses. The 153 strong
pulses are mostly within the longitude ranges of the MP and IP of the strong-mode profile,
and they are mainly distributed around the phases of $120^{\circ}$, $210^{\circ}$
and $323^{\circ}$ (top panel).}
\label{fg:profiles}
\end{figure}

Figure~\ref{fg:profiles} shows the average profiles obtained by folding
4500 successive individual pulses in the strong emission block (middle
panel) and in the long weak-mode
state (bottom panel). A point of minimum intensity throughout the strong-mode
profile is defined as both the zero of longitude and the zero of intensity.
The pulse profile of the strong mode in middle panel of Figure~\ref{fg:profiles}
basically spans $360^{\circ}$ in longitude, implying that this pulsar is most probably
an almost-aligned rotator \citep{ggks+04,elg+05,bgg+08}. The strong-mode profile of the pulsar
shows two components, i.e., main-pulse (MP) and interpulse (IP) \citep{elg+05,bgg+08}.
The phases of the two peaks of the weak-mode profile in bottom panel are coincident with that of
the leading and trailing edges of the MP. As shown in the upper panel of Figure~\ref{fg:profiles},
the 153 strong pulses are mainly distributed around the phases of $120^{\circ}$, $210^{\circ}$ and
$323^{\circ}$, which are coincident with the peak phase of the IP and the phases of the leading and
trailing edges of the MP. Most of them are within the longitude ranges of the
MP and IP of the strong-mode profile.

\begin{figure}[h,t]
\centerline{\includegraphics[angle=-90,width=0.75\textwidth]{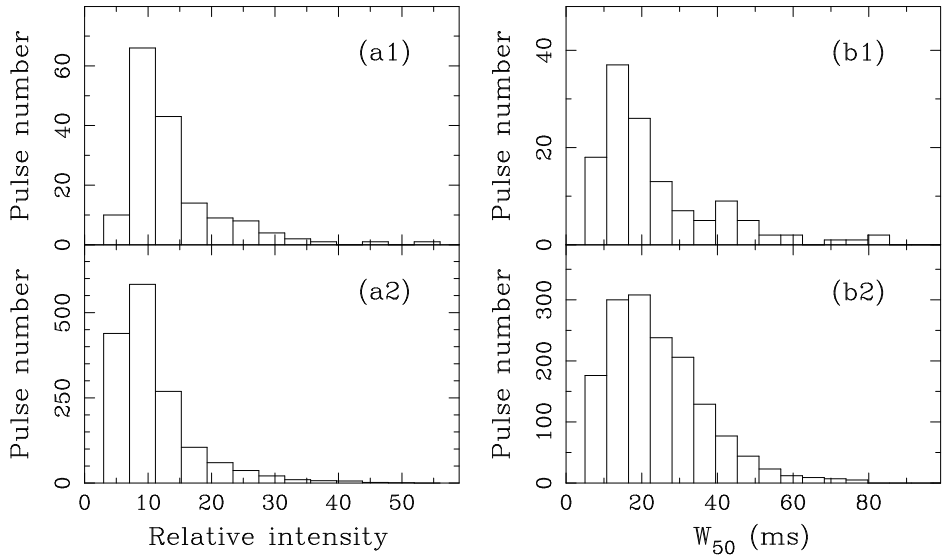}}
\caption{The histograms of relative intensity (panel a1) and W$_{50}$ (panel b1) of the 153 strong pulses
detected in the long weak-mode durations comparing with those of 1540 subpulses in the bright emission block
(panel a2 and b2), respectively.}
\label{fg:intw50}
\end{figure}

Esamdin et al. (2005) showed that up to 13 bands of drifting sub-pulses
are detected across a given pulse at 1374 MHz with as many as nine
subpulses clearly seen in individual pulses. While most strong pulses
occurring during the weak-mode state are isolated, two pulses were detected
in each of 18 individual pulses and three in each of four individual
pulses. Four pairs of pulses were detected in successive periods but
for the remainder, pulse separations ranged up to about 380 s.

Figure~\ref{fg:intw50} compares the peak flux intensities and the full width at half maximum (FWHM, W$_{50}$
hereafter) of the strong pulses detected during the weak-mode state with those of subpulses in bright emission
block respectively. These strong pulses have an average pulse width, W$_{50}$, of 24 ms,  and the peak flux intensities
of them are in the range of that of the sub-pulses in the bright emission block. The distributions of peak flux intensity and
W$_{50}$ of these pulses are essentially identical with those of the sub-pulses in the bright emission block.

\section{Periodicity of the weak-mode emission}
\citet{elg+05} detected the weak-mode average profile by folding the
individual pulses in the long ``null'' state of PSR B0826-34, and
noted that the intensity of the average profile of the weak mode is
only about 2\% of that of the strong mode. However, the identity of
the emission which contributes to form the shape of average profile of
the weak mode is not clear. Although the periodicity of the pulsar is
hardly detected through them, the possibility of these
strong pulses forming the shape of the average profile of the weak
mode should be checked.

\begin{figure}[h,t]
\centerline{\includegraphics[angle=-90,width=0.65\textwidth]{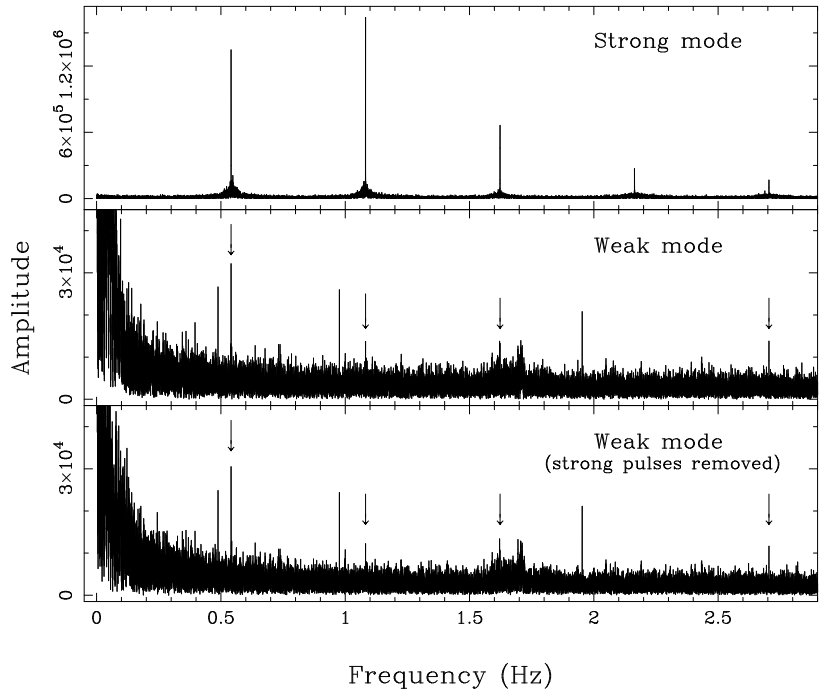}}
\caption{Spectra of the strong-mode (top panel) and the weak-mode
  emission (middle panel) of PSR B0826-34. Each spectrum is obtained
  by performing FFT to a data section of 106 min duration. The rotation frequency of the pulsar
 (0.540892 Hz) and its four successive harmonics are shown in top panel. In the
  weak-mode state, the rotation frequency and its second, third and fifth
  harmonics are detected and indicated by arrows. The spectrum of the same
  weak-mode data with the strong pulses removed is presented in the
  bottom panel.}
\label{fg:fft}
\end{figure}

The periodicity of the weak-mode emission is investigated using the
Fast Fourier Transformation (FFT). Figure~\ref{fg:fft} compares the
FFT results obtained from the strong-mode and weak-mode time
series. For the strong-mode block the rotation frequency of the pulsar
(0.540892 Hz) and its four successive harmonics (1.081785, 1.622677,
2.16349, 2.7043 Hz) are clearly shown in the upper panel of
Figure~\ref{fg:fft}. The feature of bands of drifting subpulses
are also seen in this spectrum. The rotation frequency and its second, third and fifth
harmonics (indicated by arrows in middle and bottom panels of Figure~\ref{fg:fft})
are also detected with much lower spectrum amplitudes in the weak-mode
data. The other peaks shown in the spectra in middle and bottom panels
are caused by instrumental interference. The amplitude of the rotation frequency in the weak mode is
only about 2.3\% of that in the strong mode. Although the emission is
very weak, the periodic character of the weak-mode emission is
obvious.

In bottom panel of Figure~\ref{fg:fft} we show the FFT of the same
weak-mode data section but with the strong individual pulses set to
zero amplitude. The rotation frequency and its second, third and fifth
harmonics are still seen in the spectrum. The amplitude of the fundamental
frequency is only reduced by 5\% compared to that of middle panel, indicating
that a large part of the energy of the weak-mode average profile come from
periodic weak emission.  Although the strong pulses detected in the weak-mode
state are very strong relative to the weak-mode emission, they are short and rare
and have only a minor effect on the shape of weak-mode average profile. The essential
character of the emission during the weak-mode state of the pulsar is the co-existence
of the periodic weak emission and the sporadic strong pulses.

\section{DISCUSSION}
Pulsar nulling have been detected in about 100 pulsars \citep{bac+70,hw+74,rit+76,big+92,viv+95,wmj+07,gjk+12}.
The nulls of some pulsars are very short (just a few periods), while others are
only switched on for a few per cent of the time \citep{wmj+07}. The weak-mode emission
of PSR B0826-34 is so weak that it has long been considered a null \citep{dll+79,bmh+85}.
However, the detection of the weak-mode profile, and especially the periodic weak emission
during the weak-mode of this pulsar, implies that there is not an absolute ceasing of radio
emission and supports the suggestion that nulling is a type of mode changing \citep{wmj+07}.

The non-detection of the weak-mode profile of the pulsar in GMRT data
at six frequencies \citep{bgg+08} indicates that the periodic emission
is indeed very weak. The minimum detectable flux density (5$\sigma$) for our observation
at 1374MHz is about 0.6 mJy (the details for the calculation presented by \citet{mlc+01}),
which is much lower than non-detection limit of 15 mJy of GMRT at 1060 MHz noted by \citet{bgg+08}.
However, we expect that a single-pulse search on GMRT data would reveal the strong pulses
in the weak-mode state.

The intensities and widths of these strong pulses are comparable with
that of the sub-pulses in the strong mode, and the phases of them are within the phase ranges
of the MP and IP of strong-mode profile. These may indicate that the underlying radiation mechanisms
of the strong pulses in weak-mode state and the strong-mode state are most possibly identical,
i.e., that there are very short timescale ($<1$ period) turn-on of strong-mode emission of the
pulsar during the weak-mode state.

The behavior of the sporadic bursts of strong pulses during the
weak-mode state of PSR B0826-34 seems similar to that of the
rotating radio transients (RRATs). RRATs are characterised by isolated
strong radio pulses, typically from 2 to 30 ms duration
\citep{mll+06}. A typical RRAT does not show
detectable regular emission, causing the Fourier-based searches to
fail in detecting a periodic signal. However, the timing analysis of
the sporadic pulses has confirmed that RRATs are indeed periodic
rotators, with the periods ranging from 0.4 to 7 s
\citep{mll+06,mlk+09,kkl+11}. More than 50 RRATs have been found so
far \citep{mll+06,den+09,kle+10,bsb+10,kkl+11,bbj+11}. Several pairs of pulses
have been found from RRAT J1819-1458 \citep{esam+08,hey+11}. Underlying periodic weak
emission has been detected in some RRATs \citep{bbj+11,kkl+11}. Recent
investigations show that RRATs likely represent a combination of
neutron star source populations rather than a single class \citep{kle+10,bsb+10,kkl+11}.
PSR B0826-34 presents the weak mode up to 70 \% of time. If by chance there was no
block of strong emission during an observation, the pulsar would be certainly identified as a RRAT.

As shown in upper panel of Figure~\ref{fg:profiles}, the strong pulses detected in weak-mode
state of PSR B0826-34 cover a large range of longitude, and are mainly
distributed around three phase regions. In much smaller longitude range, phase jitter
of single pulses between different phase regions have also been detected in some
RRATs \citep{esam+08, lmk+09,kkl+11}. PSR B0826-34 is a pulsar with one of the widest
known pulse profile, indicating that the pulsar is most probably a highly-aligned pulsar with the
line of sight within the radiation beam in nearly whole pulse period \citep{ggks+04,elg+05,bgg+08}.
The large phase jitter shown in Figure~\ref{fg:profiles} makes it difficult to determine the
period of the pulsar during its weak-mode state by merely timing these sporadic strong pulses.
Similarly, if the scenario of radiation-geometry of pulsars is also applicable to RRATs, the period
of a highly-aligned RRAT may also be very difficult to identify due to large phase jitter
of detected single pulses. However, highly-aligned RRATs should be very rare in RRAT population.

\citet{wsr+06} argue that PSR B0656+14 would have been classified as an RRAT were it not so nearby,
and suggest that at least some RRATs may be weak and distant pulsars
with a high modulation index. Recently, \citet{wje+11} detected two
transient average-profile components of PSR J1119-6172 showing
RRAT-like emission, also implying a link between underlying emission
mechanisms of pulsars and RRATs. \citet{bsb+10} find that PSR J0941-39 switches between RRAT-like and
pulsar-like modes, i.e. sometimes appearing with sporadic RRAT-like
emission, and at other times emitting as a bright regular nulling
pulsar. They note that this pulsar may represent a direct link between
pulsars and RRATs, and suggest that RRATs may be an evolutionary phase
of pulsars with a high nulling fraction \citet{bsb+10}. PSR B0826-34 is the second
pulsar known witches between RRAT-like and pulsar-like modes, may also be an example of such a transition from a
bright pulsar phase to a RRAT phase.

The individual pulse intensities of most radio pulsars normally
fluctuate by several times the intensity of the average pulse
\citep{rit+76,kkg+03}. However, rare and highly-modulated strong pulses are found in
three relatively old pulsars ($>$$10^{7}$ yr) B0031-07 \citep{ke+04}, B1112+50 \citep{ek+03},
J1752+2359 \citep{ek+05}. These strong pulses are unlike the giant pulses detected in some
pulsars (e.g. \citet{ss+70,hkw+03,cai+04,kbm+05}). They have different properties from the regular
weak emission of these pulsars, and cannot be explained by interstellar scintillation \citep{ek+03,ke+04,ek+05}.
Similar to the emission of PSR B0826-34 in its weak-mode state, the sporadic strong pulses of them are accompanied by
regular weak emission. These pulsars are relatively nearby pulsars and therefore the periodic weak emission is easily
detected. We suggest that the phenomena of nulling, RRATs and the extreme strong
pulses co-existing with periodic weak emission in some nearby pulsars are all manifestations of
strong and weak mode switching with very different timescales for the strong-mode persistence.

\section{CONCLUSION}

PSR B0826-34 switches between a strong mode and a weak mode with
typical timescales of hours. By analyzing a Parkes dataset obtained at
1374 MHz, we show for the first time that sporadic strong pulses
co-exist with periodic weak emission during the weak-mode state of
the pulsar. A total of 153 strong pulses, with $S/N\geq5$, are detected
during 178 minutes of weak-mode state with a few occurring within a pulse
period. The detection rate of the strong pulses is about 51.4 per hour.
These pulses have an average pulse width, W$_{50}$, of 24.0 ms. The
intensities and widths of these strong pulses are comparable with those
of the subpulses in the strong mode, and their phases are mainly distributed within
the MP and IP of the strong-mode average profile. The results indicate
that these pulses are most possibly strong-mode pulses with a very short
on-timescale.

PSR B0826-34 in its weak-mode state has properties identical to a typical RRAT.
This pulsar is the second pulsar found which oscillates between a
pulsar phase and a RRAT phase. PSR B0826-34 is a relatively old pulsar which may
be transiting from a bright phase to a weak phase. The almost-aligned rotator, the
switching between the strong and weak mode, the remarkable drifting
behavior in the strong-mode, and the periodic weak emission and strong
sporadic pulses during the weak-mode state make PSR B0826-34 an
amazing object illustrating many facets of the radio radiation mechanism of
neutron stars. Further investigations of this pulsar are likely to be fruitful.

\begin{acknowledgements}
  This work is supported by National Natural Science Foundation of
  China (NSFC) under No. 10973026 and 11273051. The Parkes radio telescope is part
  of the Australia Telescope which is funded by the Commonwealth
  Government for operation as a National Facility managed by CSIRO.
\end{acknowledgements}


\begin{thebibliography}{28}
\expandafter\ifx\csname natexlab\endcsname\relax\def\natexlab#1{#1}\fi

\bibitem[{Backer(1970)Backer}]{bac+70}
Backer, D.~C. 1970, Nature, 228, 42

\bibitem[{Biggs {et~al.}(1985)Biggs, McCulloch, Hamilton, Manchester \& Lyne}]{bmh+85}
Biggs, J.~D., McCulloch, P.~M., Hamilton, P.~A., Manchester, R.~N., \& Lyne, A.~G. 1985, MNRAS, 215, 281

\bibitem[{Biggs (1992)Biggs}]{big+92}
Biggs, J.~D. 1992, ApJ, 394, 574

\bibitem[{Bhattacharyya {et~al.}(2008)Bhattacharyya, Gupta, Gil}]{bgg+08}
Bhattacharyya, B., Gupta, Y., \& Gil, J. 2008, MNRAS, 383, 1538

\bibitem[{{Burke-Spolaor} \& {Bailes}(2010)Burke-Spolaor \& Bailes}]{bsb+10}
{Burke-Spolaor}, S., \& {Bailes}, M. 2010, MNRAS, 402, 855

\bibitem[{Burke-Spolaor {et~al.}(2011)Burke-Spolaor, Bailes, Johnston, Bates, Bhat, Burgay, D'Amico, Jameson, Keith \& Kramer}]{bbj+11}
 Burke-Spolaor, S., {et~al.} 2011, MNRAS, 416, 2465

\bibitem[{Cairns(2004)Cairns}]{cai+04}
Cairns, I.~H. 2004, ApJ, 610, 948

\bibitem[{Deneva {et~al.}(2009)]Deneva, Cordes, Mclaughlin, Nice, Lorimer, Crawford, Bhat, Camilo, Champion \& Freire}]{den+09}
Deneva, J.~S., {et~al.} 2009, ApJ, 703, 2259

\bibitem[{Durdin {et~al.}(1979)Durdin, Large, Little, Manchester, Lyne \& Taylor}]{dll+79}
Durdin, J.~M., Large, M.~I, Little, A.~G, Manchester, R.~N., Lyne A.~G., \& Taylor, J.~H. 1979, MNRAS, 186, 39p

\bibitem[{{Ershov} \& {Kuzmin}(2003)Ershov \& Kuzmin}]{ek+03}
Ershov, A.~A., \& Kuzmin, A.~D. 2003, AstL, 29, 91

\bibitem[{{Ershov} \& {Kuzmin}(2005)Ershov \& Kuzmin}]{ek+05}
Ershov, A.~A., \& Kuzmin, A.~D. 2005, A\&A, 443, 593

\bibitem[{Esamdin {et~al.}(2005)Esamdin, Lyne, Graham-Smith, Kramer, Manchester \& Wu}]{elg+05}
Esamdin, A., Lyne, A.~G., Graham-Smith, F., Kramer, M., Manchester,R.~N., \& Wu, X. 2005, MNRAS, 356, 59

\bibitem[{Esamdin {et~al.}(2008)Esamdin, Zhao, Yan, Wang, Nizamdidin}]{esam+08}
Esamdin A., Zhao C.~S., Yan Y., Wang N., Nizamidin H., \& Liu Z.~Y. 2008, MNRAS, 389, 1399

\bibitem[{Gajjar {et~al.}(2012)Gajjar, Joshi \& Kramer}]{gjk+12}
Gajjar, V., Joshi, B.~C., \& Kramer, M. 2012, MNRAS, 424, 1197

\bibitem[{Gupta {et~al.}(2004)Gupta, Gil, Kijak \& Sendyk}]{ggks+04}
Gupta, Y., Gil, J., Kijak, J., \& Sendyk, M. 2004, A\&A, 426, 229

\bibitem[{Hankins {et~al.}(2005)Hankins, Kern, Weatherall \& Eilek}]{hkw+03}
Hankins T. H., Kern J. S., Weatherall J. C., \& Eilek J. A. 2003, Nat, 422, 141

\bibitem[{{Hesse} \& {Wielebinski}(1974)Hesse \& Wielebinski}]{hw+74}
Hesse, K.~H., \& Wielebinski, R. 1974, A\&A, 31, 409

\bibitem[{Hu {et~al.}(2011)Hu, Esamdin, Yuan, Liu, Xu, Li, Tao \& Wang}]{hey+11}
Hu H. D., Esamdin A., Yuan J.~P., Liu Z.~Y., Xu R.~X., Li J., Tao G.~C., \& Wang N. 2011, A\&A, 530, 67

\bibitem[{Keane {et~al.}(2010)Keane, Ludovici, Eatough, Kramer, Lyne, McLaughlin, Stappers}]{kle+10}
Keane, E.~F., Ludovici, D.~A., Eatough, R.~P., Kramer, M., Lyne, A.~G., McLaughlin, M.~A., \& Stappers, B.~W. 2010, MNRAS, 401, 1057

\bibitem[{Keane {et~al.}(2011)Keane, Kramer, Lyne, Stappers \& McLaughlin}]{kkl+11}
Keane, E.~F., Kramer, M, Lyne, A.~G., Stappers, B.~W., \& McLaughlin, M.~A. 2011, MNRAS, 415, 3065

\bibitem[{Knight {et~al.}(2005)Knight, Bailes, Manchester}]{kbm+05}
Knight H. S., Bailes M., Manchester R. N., et al. 2005, ApJ, 625, 951

\bibitem[{kramer {et~al.}(2003)Kramer, Karastergiou, Gupta}]{kkg+03}
Kramer, M., Karastergiou, A., Gupta, Y., et al. 2003, A\&A, 407, 655

\bibitem[{{Kuzmin} \& {Ershov}(2004)Kuzmin \& Ershov}]{ke+04}
Kuzmin, A.~D., \& Ershov, A.~A. 2004, A\&A, 427, 575

\bibitem[{Lyne {et~al.}(2009)Lyne, McLaughlin, Keane, Kramer, Espinoza, Stappers, Palliyaguru, Miller}]{lmk+09}
Lyne, A.~G., {et~al.} 2009, MNRAS, 400, 1439

\bibitem[{Manchester {et~al.}(1978)Manchester, Lyne, Taylor, Durdin, Large \& Little}]{mlt+78}
Manchester, R.~N., Lyne, A.~G., Taylor, J.~H., Durdin, J.~M., Large, M.~I., \& Little, A.~G. 1978, MNRAS, 185, 409

\bibitem[{Manchester {et~al.}(2001)Manchester, Lyne, Camilo, Bell, Kaspi, D'Amico, McKay, Crawford, Stairs \& Possenti}]{mlc+01}
Manchester, R.~N., {et~al.} 2001, MNRAS, 328, 17

\bibitem[{Mclaughlin {et~al.}(2006)]Mclaughlin, Lyne, Lorimer, Faulkner, Manchester \& Cordes}]{mll+06}
Mclaughlin, M.~A., {et~al.} 2006, Nature, 439, 817

\bibitem[{Mclaughlin {et~al.}(2009)]Mclaughlin, Lyne, Keane, kramer, Miller, Lorimer, Manchester, Camilo \& Stairs}]{mlk+09}
Mclaughlin, M.~A., {et~al.} 2009, MNRAS, 400, 1431

\bibitem[{Ritchings(1976)Ritchings}]{rit+76}
Ritchings, R.~T. 1976, MNRAS, 176, 249

\bibitem[{{Serylak}(2011)}]{ser+11}
{Serylak}, M. 2011, PhD thesis, University of Amsterdam, {\sf http://dare.uva.nl/en/record/369352}

\bibitem[{{Staelin} \& {Sutton}(1970)Staelin \& Sutton}]{ss+70}
Staelin, D.~H., \& Sutton, J.~M. 1970, Nature, 226, 69

\bibitem[{{van Leeuwen} \& {Timokhin}(2012)}]{vlt+12}
{van Leeuwen}, J., \& {Timokhin}, A.~N. 2012, ApJ, 752, 155

\bibitem[{Vivekanand(1995)Vivekanand}]{viv+95}
Vivekanand M. 1995, MNRAS, 274, 785

\bibitem[{Wang {et~al.}(2007)Wang, Manchester \& Johnston}]{wmj+07}
Wang, N., Manchester, R.~N., \& Johnston, S. 2007, MNRAS, 377, 1383

\bibitem[{Weltevrede {et~al.}(2011)Weltevrede, Johnston \& Espinoza}]{wje+11}
Weltevrede, P., Johnston, S., \& Espinoza, C.~M. 2011, MNRAS, 411, 1917

\bibitem[{Weltevrede {et~al.}(2006) Weltevrede, Stappers, Rankin \& Wright}]{wsr+06}
Weltevrede, P., Stappers, B.~W., Rankin, J.~M., \& Wright, G.~A.~E. 2006, ApJ, 645, L149


\end{thebibliography}

\end{document}